\documentclass{ws-mpla}

\begin{document}

\markboth{J.G. Messchendorp}
{Few-nucleon studies at intermediate energies}

\catchline{}{}{}{}{}

\title{FEW-NUCLEON STUDIES AT INTERMEDIATE ENERGIES}

\author{\footnotesize J.G. Messchendorp}

\address{University of Groningen, Kernfysisch Versneller Instituut, Zernikelaan 25, 9747 AA, Groningen, 
The Netherlands.\\ messchendorp@kvi.nl}

\maketitle

\pub{Received (Day Month Year)}{Revised (Day Month Year)}

\begin{abstract}

Observables in proton-deuteron scattering are sensitive probes
of the nucleon-nucleon interaction and three-nucleon force
effects. Several facilities, including the KVI, allow a detailed study of 
few-nucleon interactions below the pion-production threshold
exploiting polarized proton and deuteron beams.
In this contribution, some recent results 
are discussed and interpreted exploiting rigorous Faddeev calculations.
Furthermore, an experimental inconsistencies between two measurements
of the cross section in elastic proton-deuteron scattering 
at 135~MeV/nucleon is reviewed.

\end{abstract}

\ccode{PACS Nos.: 21.30.-x, 21.45.-v, 24.70.+s, 25.45.De}

\section{Introduction}	

The nucleon-nucleon potential (NNP) has been studied extensively by
investigating the properties of bound nuclear systems and, in more
detail, via a comparison of high-precision two-nucleon scattering data
with modern potentials based on the exchange of
bosons~{\cite{bon87,argon,nij1}}. A few of the modern NNPs were
facilitated by a partial-wave analysis, which provides a nearly
model-independent analysis of the available scattering
data~{\cite{nij2}}. The modern NNPs reproduce the world database with
a reduced chi-square close to one and have, therefore, been accepted
as high-quality benchmark potentials. The precision of modern NNPs
has given scientists the confidence to study in detail the
three-nucleon potential (3NP) that was already predicted in 1939 by
Primakoff {\it et al.}~{\cite{prima}}. Compelling evidence of 3NP effects
has come from various recent theoretical and experimental studies. For
example, for light nuclei, Green's function Monte Carlo calculations
employing the high-quality NNPs clearly underestimate the experimental
binding energies~{\cite{argon}} and, therefore, show that NNPs are not
sufficient to describe the three-nucleon and heavier systems
accurately. Deficiencies of theoretical predictions based on pair-wise 
nucleon-nucleon potentials have been observed in three-nucleon scattering 
observables as well. 

Most of the present-day 3NPs are based on a refined version of the Fujita-Miyazawa 
force~{\cite{fuji}} in which a 2$\pi$-exchange 
mechanism is incorporated by an intermediate $\Delta$ excitation of one of the nucleons. 
Later, more refined ingredients have been added as in Urbana IX~{\cite{urbana}} and 
Tucson-Melbourne (TM')~{\cite{tuscon}} allowing for additional processes contributing 
to the rescattering of the mesons on an intermediate excited nucleon. A different 
approach is provided by the Hannover theory group,  where the $\Delta$-isobar is treated 
on the same basis as the nucleon, resulting in a coupled-channel potential 
CD-Bonn+$\Delta$~{\cite{hannover}} with pair-wise nucleon-nucleon and nucleon-$\Delta$ 
interactions mediated through the exchange of $\pi$, $\rho$, $\omega$, and $\sigma$ mesons. 
Within this self consistent framework, the $\Delta$-isobar excitation mediates an 
effective 3NP with prominent Fujita-Miyazawa and Illinois ring-type contributions
~{\cite{pieper}}.

One of the experimental programs at KVI focuses on obtaining high-precision 
data in the few-nucleon scattering processes below the pion-production threshold. 
The goal is to study the details of the nucleon-nucleon and three-nucleon interactions 
through a comparison with predictions from state-of-the-art effective nucleon-nucleon 
potentials and models based on a chiral-symmetry expansion. 
For this purpose, cross sections and analyzing powers are measured in few-nucleon scattering
processes. The focus of the few-body program at KVI is mainly oriented towards understanding 
three- and, more recently, four-nucleon systems by exploring $p+d$, $d+p$, and $d+d$ reactions 
with polarized proton and deuteron beams. Different final states have been observed which includes the 
elastic, break-up, and radiative capture reactions. In this paper, some recent results for the 
elastic proton-deuteron scattering are presented. Elsewhere in these proceedings,
preliminary results for the proton-deuteron break-up channel, M. Eslami-Kalantari {\it et al.}, 
and elastic deuteron-deuteron channel, A. Ramazani-Moghaddam-Arani {\it et al.}, are discussed.

\section{Recent results and observations in elastic proton-deuteron scattering}

\begin{figure}[ht]
\centerline{\psfig{file=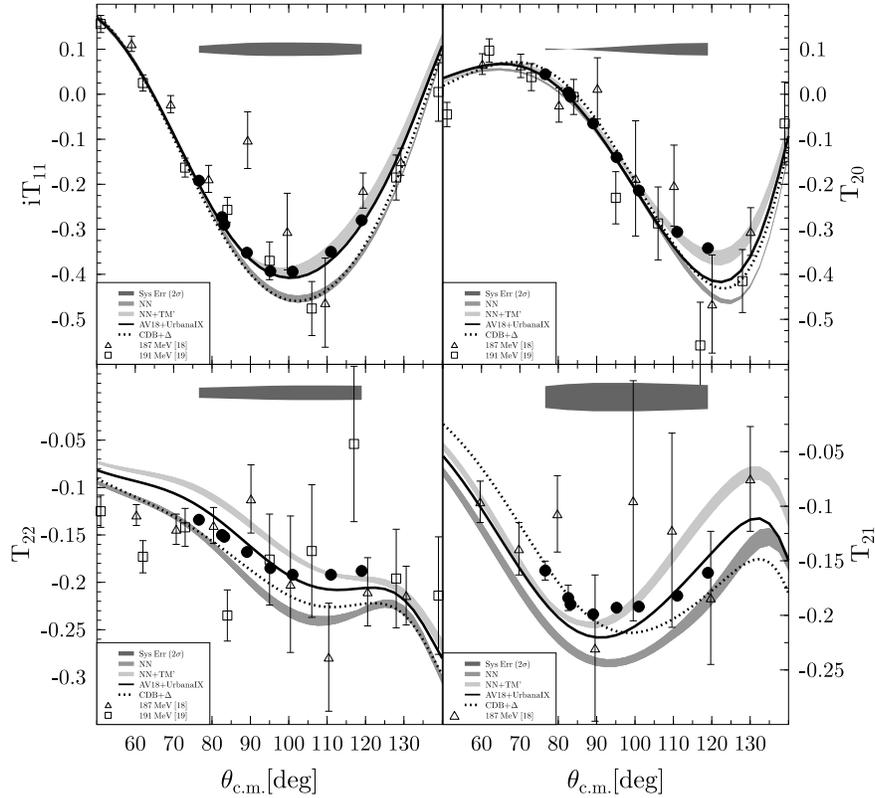,width=4.5in}}
\vspace*{-8pt}
\caption{Vector and tensor analyzing powers in the elastic $\vec d+p$ scattering at an 
incident deuteron beam energy of $E^{lab}_{d}$=180~MeV. The open triangles are data from 
Ref.~\protect\cite{witala} and the open squares are data from Ref.~\protect\cite{garcon}.
The dark gray bands at the top of the panels represent the systematical uncertainty 
($2\sigma$) for every data point. The other dark gray bands correspond to
calculations including only two-nucleon potentials. The light gray bands represent 
calculations including an  additional Tucson-Melbourne TM' three-nucleon force as well. 
The solid lines correspond to results of a Faddeev calculation using the AV18 two-nucleon 
potential combined with the Urbana-IX (UIX) three-nucleon potential. The dotted lines represent 
the results of a coupled-channel potential CDB+$\Delta$ calculations.
\protect\label{fig1}}
\end{figure}

In the last decade, high-precision data at intermediate energies in 
elastic $Nd$ and $dN$ scattering for a large energy range 
together with rigorous Faddeev calculations for the three-nucleon system have
proven to be a sensitive tool to study the 3NP. In particular, a large
sensitivity to 3NP effects exists in the minimum of the differential
cross section. Precision data for a large energy interval for the differential 
cross section and analyzing power have come from recent experimental studies 
at KVI~{\cite{kars01,kars03,kars05}}, Research Center for Nuclear Physics 
(RIKEN)~{\cite{sakai00}} and RCNP~{\cite{{kimiko05}}. 

A comparison between data and results from Faddeev calculations showed that
our present understanding of the 3NP is not sufficient to describe all the observables
in the elastic channel. Even at relatively low energies, $\approx$100~MeV/nucleon, 
significant discrepancies appear, in particular in the tensor-analyzing powers.
Figure~\ref{fig1} shows the results of a measurement of the vector 
and tensor analyzing powers in the elastic deuteron proton scattering process~\cite{mar07}. 
The data were obtained at the RIKEN facility in Japan using a polarized deuteron beam
and were analyzed by the experimental nuclear-physics group at the KVI.
Note that the data represented by the filled circles are
of superb precision with respect to data taken earlier for these observables.
Furthermore, rigorous Faddeev calculations incorporating state-of-the-art
NNPs cannot describe the tensor observables $T_{22}$ and $T_{21}$.
The inclusion of three-nucleon force effects does not remedy the deficiency either.

\begin{figure}[ht]
\centerline{\psfig{file=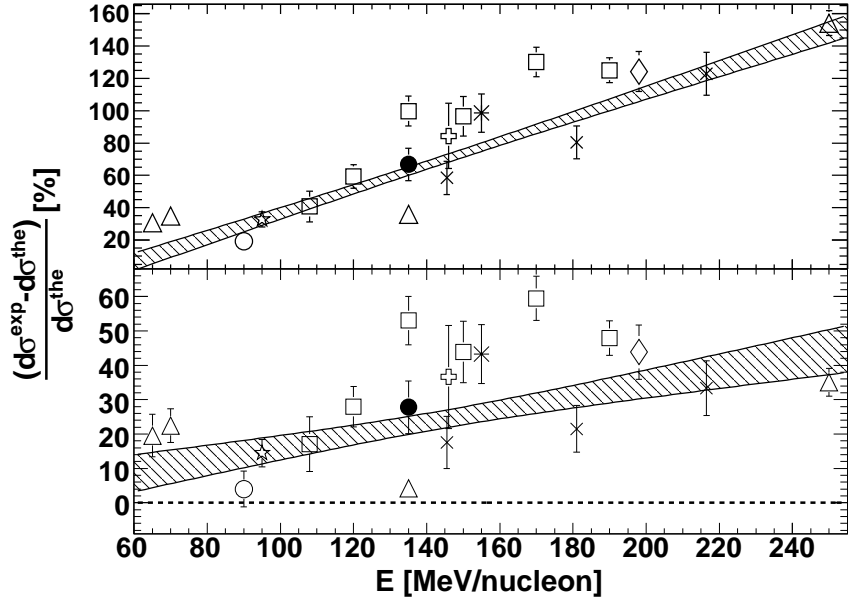,width=5.0in}}
\vspace*{-8pt}
\caption{The relative difference between the calculations by the Hannover-Lisbon theory 
group and the measured cross sections for the elastic $p+d$ reaction as a
function of beam energy for a center-of-mass angle, $\theta_{\rm c.m.}=140 ^\circ$. 
The top panel shows the differences with a calculation based
on the CD-Bonn potential and the Coulomb interaction, whereas for the 
bottom panel an additional $\Delta$ isobar has been taken into account.
Open squares are data from Ref.~\protect\cite{kars03}, open triangles are data from
Refs.~\protect\cite{sakai00,kimiko02,shimi,hatan}, open circle is from~\protect\cite{hamid},
open star is from~\protect\cite{mermod}, crosses are from~\protect\cite{Igo}, star
is from~\protect\cite{kurodo}, open cross is from~\protect\cite{postma}, diamond is from
\protect\cite{ald} and the filled circle is from~\protect\cite{ram08}. The shaded
band represents the result of a line fit through the data excluding the results obtained at 
KVI, RIKEN and RCNP. The width of the band corresponds to a 2$\sigma$ error of the 
fit.\protect\label{fig2}}
\end{figure}

The differential cross sections at relatively low energies, 50$-$100~MeV/nucleon, are 
described reasonably well by calculations based on our present understanding of NNPs and 3NPs. 
Towards larger incident energies, the contribution of the 3NPs increases dramatically.
This is illustrated in the top panel of Fig.~\ref{fig2}, which depicts the relative 
difference between various experimental cross sections taken at a center-of-mass angle, $\theta_{c.m.}$=140$^\circ$ 
and a  corresponding calculation based on a CD-Bonn NNP by the Hannover-Lisbon theory group.
The discrepancy between data and calculation at 200~MeV/nucleon is more than 100\%.
A large part of this deficiency can be remedied by the inclusion of a $\Delta$ isobar,
as a model for the 3NP and illustrated in the bottom panel of Fig.~\ref{fig2}. 
Note, however, that a significant deviation in the order of
30\% at 200~MeV/nucleon remains. Discussions are ongoing whether these discrepancies
are due to short-range 3NP or relativistic effects which are not completely or consistently
accounted for in the present models.

In the past years, a discussion was initiated within the nuclear physics community 
on the reliability of the experimental data taken at an incident energy of 135~MeV/nucleon.
At this energy, the differential cross-section data obtained at KVI were found to be
significantly larger than those measured at RIKEN and at RCNP, as can be 
observed in Fig.~\ref{fig2}. The KVI data (open squares) deviate significantly from 
predictions of state-of-the-art Faddeev calculations incorporating modern 
NNPs and 3NPs at this energy, whereas the results obtained at RIKEN and 
RCNP (open triangles) imply that the cross section can be described reasonably well 
exploiting the same potentials. A fit through the world database revealed that the 
data taken at RIKEN and at KVI deviate 8$\sigma$ and 3.5$\sigma$ from the expected trend,
respectively. The trend is shown as shaded band with its width corresponding to
a 2$\sigma$ error of the fit. Note that a more recent measurement
taken at KVI using the Big Instrument for Polarization Analysis, fits well the expected trend (filled circles). 
Similar analyses have been carried out at different center-of-mass angles. From this, we concluded that 
the previously-published data taken at KVI and RIKEN/RCNP suffered likely from an overall 
normalization problem.

\section{Summary and conclusions}

The three-nucleon scattering process at intermediate energies has demonstrated to
be a sensitive tool to study the details of few-nucleon forces. Accurate
experimental data have grown rapidly in the last decade, thereby, revealing
many new insights in the few-nucleon system. Furthermore, the calculations presently 
on the market are of very high quality: ab-initio, self-consistent, and with the ability to 
include effects such as Coulomb and relativity. 

In spite of the impressive accuracy in the data as well as in the calculation, there are still
several discrepancies to be understood in the three-nucleon scattering process at
intermediate energies. In particular, large deficiencies have been observed in polarization
observables and in the differential cross section at relatively high energies.
Part of these discrepancies might be due to short-range three-nucleon force effects
or relativistic effects which have not been taken into account so far.

Although the statistical uncertainty of the experimental data has meanwhile reached
sufficient precision for most of the observables, there remains a large uncertainty
due to systematic effects. The latter uncertainty is - in general - difficult to estimate
and can easily result in inconsistencies between data sets
taken at different laboratories. It is of crucial importance that these inconsistencies
are taken seriously. The most generic approach would be to perform a partial wave analysis, 
albeit not completely model independent, of all available three-nucleon scattering data
similar to what has been done for the nucleon-nucleon scattering data by the 
Nij\-me\-gen group~{\cite{nij2}}. 

\section*{Acknowledgments}

The author acknowledges Mohammad Eslami-Kalantari, 
Hossein Mardanpour, and Ahmad Ramazani. The results presented here
are part of their PhD theses. Furthermore, the author thanks 
Nasser Kalantar for the valuable discussions and his input.
This work is part of the research program of the ``Stichting voor
Fundamenteel Onderzoek der Materie'' (FOM) with financial support from
the ``Nederlandse Organisatie voor Wetenschappelijk Onderzoek'' (NWO).
Furthermore, the present work has been performed with financial
support from the University of Groningen and the Gesellschaft f\"ur
Schwerionenforschung mbH (GSI), Helmholtzzentrum f\"ur Schwer\-ionen\-forschung GmbH, 
Darmstadt.

\end{document}